\documentclass[11pt,twoside]{article}
\usepackage{asp2010}

\resetcounters

\aspvolume{TBA} 
\aspcpryear{2012} 
\aspvoltitle{FOUR DECADES OF RESEARCH ON MASSIVE STARS: 
A SCIENTIFIC MEETING IN HONOUR OF ANTHONY F.J. MOFFAT} 
\aspvolauthor{Laurent Drissen, Nicole St-Louis, Carmelle Robert, and 
Anthony F.J. Moffat, eds.} 

\begin{document}

\title{Spectral Identification of New Galactic cLBV and WR Stars}
\author{G.S.~Stringfellow$^1$, V.V.~Gvaramadze$^2$, Y.~Beletsky$^3$ and A.Y.~Kniazev$^{4,2}$
\affil{$^1$Center for Astrophysics and Space Astronomy, University
of Colorado, 389 UCB, Boulder, CO 80309-0389, USA}
\affil{$^2$Sternberg Astronomical Institute, Moscow State
University, Universitetskij Pr. 13, Moscow 119992, Russia}
\affil{$^3$European Southern Observatory, Alonso de Cordova 3107,
Santiago, Chile}
\affil{$^4$South African Astronomical Observatory
and Southern African Large Telescope Foundation, PO Box 9, 7935
Observatory, Cape Town, South Africa}}

\begin{abstract}
We have undertaken a near-IR spectral survey of
stars associated with compact nebulae recently revealed by the
{\it Spitzer} and WISE imaging surveys. These circumstellar
nebulae, produced by massive evolved stars, display a variety of
symmetries and shapes and are often only evident at mid-IR
wavelengths. Stars associated with $\sim$50 of these nebulae have
been observed. We also obtained recent spectra of previously
confirmed (known) luminous blue variables (LBVs) and candidate
LBVs (cLBVs). The spectral similarity of the stars observed when
compared directly to known LBVs and Wolf-Rayet (WR) stars
indicate many are newly identified cLBVs, with a few being newly
discovered WR stars, mostly of WN8-9h spectral type. These results
suggest that a large population of previously unidentified cLBVs
and related transitional stars reside in the Galaxy and confirm
that circumstellar nebulae are inherent to most (c)LBVs.
\end{abstract}

\subsection*{Nebulae as signposts to the most massive stars}

\begin{figure}[!ht]
\plotone{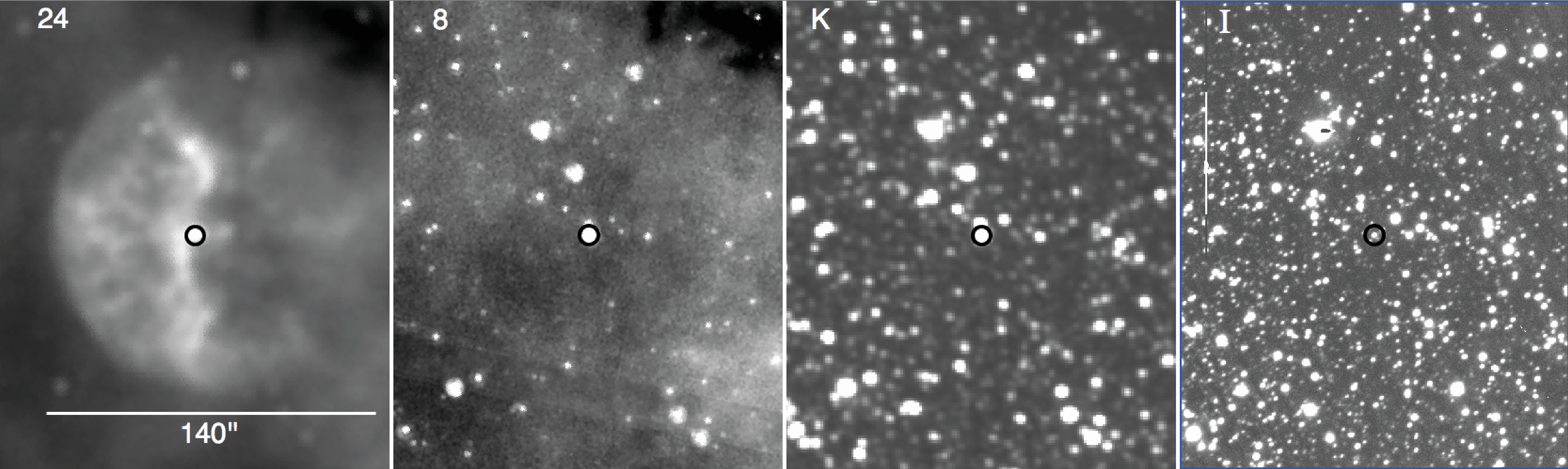}
\caption{{\it Left to right}: 24\,$\mu$m MIPS nebula MN\,96 
\citep{gv10a} with the off-center progenitor star marked, followed
by IRAC 8\,$\mu$m, 2MASS $K_{\rm s}$, and our deep FTS $I$-band
image. The spectrum of the marked star is discussed in
\citet{st12} and \citet{wa11}.} \label{fig1}
\end{figure}

The {\it Spitzer} MIPSGAL \citep{ca09} and GLIMPSE \citep{be03}
surveys, and to a lesser degree other {\it Spitzer} Legacy
Programs, have led to a new discovery path for the most massive
stars in our Galaxy by the signpost their 24 and 8\,$\mu$m bright
nebulae signal in the images. These impressive nebulae indicate
that massive star progenitors reside within, and hence where to
search for them. Optical compact circumstellar nebulae associated with
LBVs and WR stars have been known for decades
\citep[e.g.,][]{no95,do94}. The discovery potential in the IR was
demonstrated by a few examples found in the {\it MSX} mid-IR
survey data \citep[e.g.,][]{eg02,cl03}. The application of this
technique has exploded recently with huge number of such nebulae
having been identified through exploration of the extensive
archival {\it Spitzer} database
\citep*[e.g.,][]{gv10a,wa10,mi10}.
The hunt for the progenitor stars that produce these nebulae
requires spectroscopy to identify the stars and their evolutionary
state, primarily in the IR as the stars are typically undetected
in the optical. There are often at least several potential stellar
candidates. Figure \ref{fig1} illustrates the case for MN\,96  \citep[\#54 
in][]{wa10}. The
left panel displays the 24\,$\mu$m shell, with a bright source
marked to the left of center. Following this source to shorter
wavelengths to the right the 8\,$\mu$m, 2MASS $K_{\rm s}$, and our
deep $I$-band imaging all detect the source. The POSS II IR and
red plates do not clearly detect the source, whereas we have
optically recovered it. We are carrying out a spectral survey of
stars associated with the newly discovered nebulae in order to
identify and classify them. Highlights of our results are
presented here.

\subsection*{Spectroscopic identification of the progenitor stars}

\begin{figure}[!ht]
\plottwo{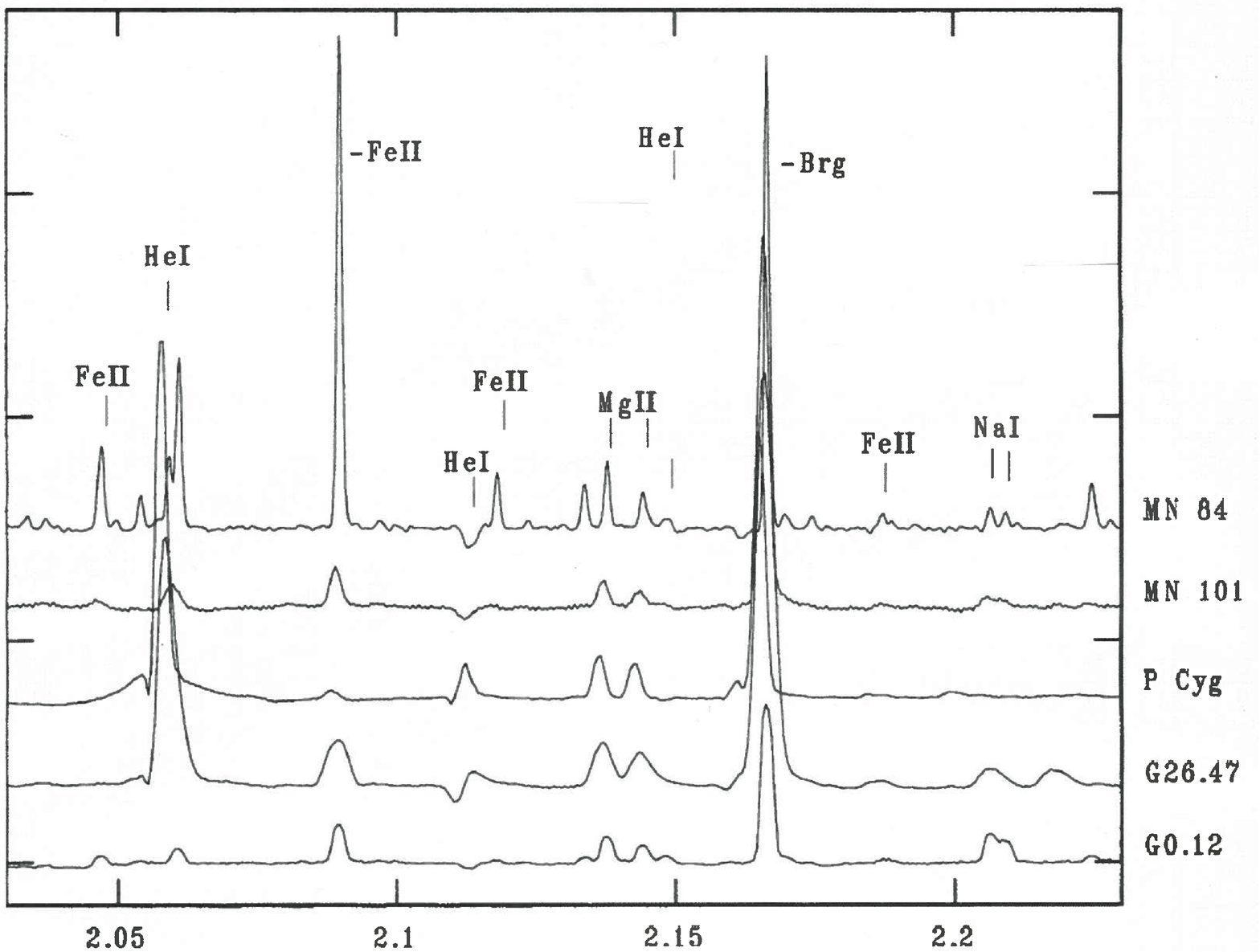}{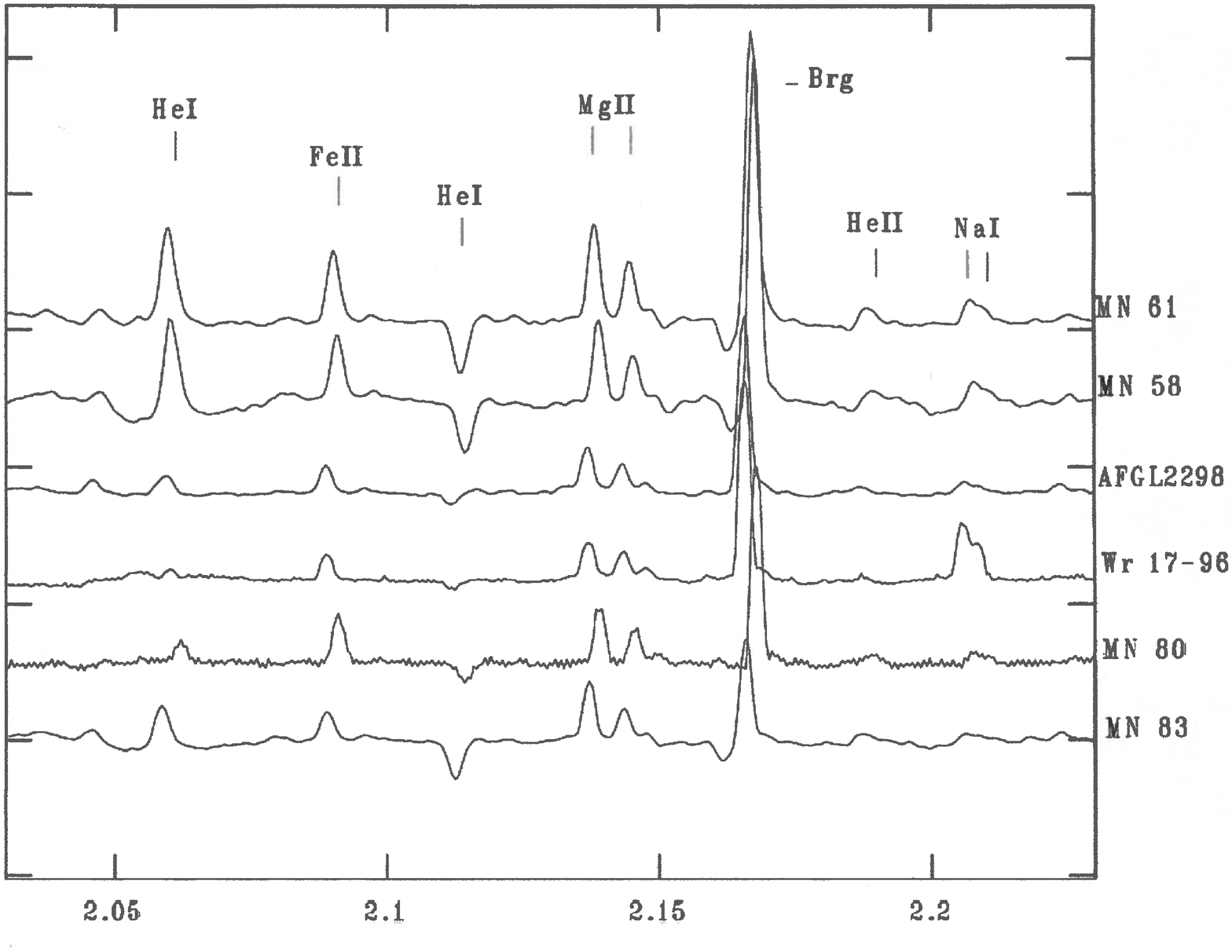}
\caption{Normalized $K$-band spectra of the central stars
associated with 6 new MIPS nebulae \citep{gv10a}. New spectra of previously known
LBVs P\,Cyg, AFGL\,2298, G0.120$-$0.048 \citep{ma10}, and cLBVs
Gal\,026.47+00.02 and Wray\,17-96 are also shown.} \label{fig2}
\end{figure}

A near-IR spectral survey of the stars
identified within the nebulae, as outlined above, has been
conducted \citep[][and in preparation]{st12}. 
Primary facilities utilized include the APO, ESO-VLT, NASA-IRTF, and
Palomar-Hale telescopes.  IR spectra have been obtained for stars associated with
$\sim$50 shells, and many appear to be cLBVs, displaying strong
emission lines of H, He\,{\sc i}, Mg\,{\sc ii}, Na\,{\sc i}, and
often, but not always, Fe\,{\sc ii}. Occasionally, [Fe\,{\sc ii}]
emission lines are present. Figure \ref{fig2} presents a selection of 6
newly identified cLBVs, along with new spectra of previously known
LBVs and cLBVs for comparison. All spectra shown in Figure \ref{fig2}
display Fe\,{\sc
ii} 2.089\,$\mu$m emission with varying strength. Most LBVs are
believed to undergo S\,Dor cycles, where their atmospheres expand
and contract, resulting in cool temperatures with large radii at
peak light, and hotter temperatures at minimum radii and
brightness \citep[e.g.,][ for the LBV AG\,Car]{st01}. If all these
stars are undergoing active S\,Dor cycles, they would correspond to
brighter phases when they are at larger radii where Fe\,{\sc ii} forms at
cooler temperatures in the cycle; monitoring of these stars
continues. 

\citet{st12} present a few examples of cLBVs from our survey 
that are void of Fe\,{\sc ii} emission in their $K$-band spectra, including 
MN\,96, MN\,112, and MN\,76.  The near-IR spectra of these stars are classified as 
Fe\,{\sc ii} deficient cLBVs, perhaps indicating they are currently in
a hot S\,Dor phase, and/or transitioning to a late-WN phase.  
While Fe\,{\sc ii} 2.089\,$\mu$m emission 
is weak in P\,Cygni compared to many other (c)LBVs, and relative to its own 
emission lines in the $K$-band (see Figure \ref{fig2}), Fe\,{\sc ii} is entirely absent 
in the $K$-band spectra presented in \citet{st12}, including MN\,112. However, 
MN\,112 is virtually 
identical to P\,Cygni over the optical spectral range shown by \citet{gv10b}, 
which contains numerous Fe\,{\sc iii} lines but no Fe\,{\sc ii} lines,
indicating a hotter line forming region. 
The star associated with MN\,80 was previously classified as G7\,I \citep{wa10}, 
which is clearly disparate with the spectrum shown in Figure \ref{fig2}.
There are two possible reasons for this: either a different star
was observed by \citet{wa10}, or the spectrum has changed dramatically between the
two observations. The complete survey has identified dozens of new
cLBVs, considerably increasing the known Galactic population of
such stars, and additional WR stars \citep[see also][and in preparation]{st12}.

\subsection*{Discovery space of mid-IR shells remains vast}

The MIPS nebulae catalogues on which our spectroscopic study is
based remain incomplete. We have discovered further
nebulae \citep[e.g.,][]{gv11} using the WISE all-sky survey 
\citep{wr10}. Optical
follow-up spectroscopy of two central stars with SALT indicate
their spectra are very similar to those of the prototype LBV
P\,Cygni, which implies the LBV classification for these stars as
well (Gvaramadze et al., in preparation). Our study continues with
further discoveries anticipated.

\acknowledgements GSS thanks the AAS for awarding a Small Research
Grant for travel support for the observing runs and partial
support to attend this meeting.

\end{document}